Review

*Development and application of SEM/EDS in biological, biomedical & nanotechnological research*

Aniruddha Acharya[1]✉

1 Division of Mathematics & Sciences; Delta State University

✉ aniruddha1302@gmail.com & aacharya@deltastate.edu

**Abstract**

This comprehensive review discusses the development of scanning electron microscopy and the application of this technology in different fields such as biology, nanobiotechnology and biomedical science. Besides being a tool for high resolution imaging of surface or topography, the technology is coupled with analytical techniques such as energy dispersive spectroscopy for elemental mapping. Since the commercialization of the technology, it has developed manifold and currently very high-resolution nano scale imaging is possible by this technology. The development of FIB-SEM has allowed three-dimensional imaging of materials while the development of cryo-stage allows imaging of hydrated biological samples. Though variable pressure or environmental SEM can be used for imaging hydrated samples, they cannot capture a high-resolution image. SBEM and ATUM-SEM has automated the sampling process while improved and more powerful software along with user-friendly computer interface has made image analysis faster and more reliable. This review presents one of the most widely used analytical techniques used across the globe for scientific investigation. The power and potential of SEM is expanding with the development of accessory technology.

**Keywords:** Scanning electron microscope, energy dispersive spectroscopy, nanobiotechnology, biological science, biomedical science

**Development of SEM**

Since the development of scanning electron microscopy, a significant improvement in the instrument components such as electron gun, electromagnetic lenses, detectors, camera, and sample preparation protocols such as immunolabeling, critical point drying, freeze fracture technique and, high-pressure freezing has resulted in high resolution imaging capabilities (Table 1). The advanced and most recent scanning electron microscopes have sub-nanometer resolution and are widely used in biological science, biomedical science, and material science investigations (Hosoya et al. 2019). Besides the tungsten filament which is still very common, the development of lanthanum hexaboride filament and field emission gun generated a brighter source of electron beam and significant improvement were possible to reduce the size or diameter of the electron beam (probe) that was used to scan the sample surface. Such improvements allowed the use of beams generated by low accelerating voltage thus reducing noise and improving the imaging of finer details of surfaces. Improvement in electrical conductivity of non-conductive biological sample by coating them with metals such as gold, palladium, iridium, tungsten or platinum reduced charging and improved resolution and image quality of non-conductive biological samples (de Souza & Attias 2018).

**Volume microscopy**

Volume microscopy is a relatively new field of technology where highly reliable microscopic data can be obtained at a nanometer-scale. This technique can image materials, cells, and tissues in three dimensions and is increasingly being used in biological and material science research. The imaging technique is automated, software-driven, and significantly faster than the traditional imaging techniques using transmission electron microscope which are error-prone and cumbersome. Volume microscopy generates a large volume of data through serial image acquisition, has high z-resolution and thus is more convenient than manual sample preparation and sectioning that has large chances or error or artifacts. Serial images can be procured without the need of handling large number of sections and imaging them one at a time using transmission electron microscopy. The 3D datasets obtained by volume microscopy can be analyzed by associated software and can be utilized for qualitative and quantitative data extraction. (Titze & Genoud 2016).

**Types of scanning electron microscopy**

Since the development of SEM in the 1960s and their subsequent commercialization by Cambridge Scientific Equipment in 1965 they have gone through significant improvement in terms of their versatility and resolution limits. During the early years of SEM development, they had a resolution limit of 50 nm, however, currently the resolution limit of 3 nanometer or less is very common. The Helium Ion Microscopy (HIM), Focused Ion Beam (FIB), Microtome Serial Sectioning (MSS), environmental or low vacuum scanning microscopy and cryo-scanning electron microscopy had added significant amount of versatility to the traditional scanning electron microscopy or scanning microscopy and has allowed procurement of significant amount of cellular data. Simultaneous improvement in software has made it possible to process significant amount of microscopic data in a relatively short time (Bogner et al., 2007).

**Automated tape-collecting ultramicrotome SEM (ATUM-SEM)**.

ATUM-SEM or array tomography scanning electron microscopy is an important technique for volume microscopy. This technique combines the capabilities of array tomography and scanning electron microscopy. In array tomography sections from a sample can be made manually which can be collected serially on glass slides. The purpose of array tomography was to comparatively image samples both using light microscopy and electron microscopy to have a deeper understanding of structural and functional relationship within a cell. Since electron microscopy can only depict information about structure, it was necessary to use light or laser microscopy assisted with fluorescent dyes to understand the functional component that is related to cell architecture. In ATUM-SEM the sample is mounted in an ultramicrotome that is equipped with diamond knife. The diamond knife is used for sectioning and thin sections of approximately 50 nm thickness can be done which float in a water boat. The water in the boat can be replaced and the sections can be collected in a tape controlled by a conveyer belt mechanism for further processing. The tape can collect thousands of sections that can easily get attached to the tape. The tape can be rolled to a reel and in such a way the samples can be preserved for a long time. This method of sample collection and preservation is more stable than compared to the samples that are collected in a grid and stored in grid box for transmission electron microscopy imaging. The only drawback of this method is that the samples cannot be used for transmission electron microscopy imaging as their support matrix is not electron transparent. All the process described above from sectioning to collection is done automatically while the imaging using the scanning electron microscope is done semi-automatically. ATUM-SEM is an important technology used in cell biology to understand cellular functions (Titze & Genoud 2016; Laws et al. 2022).

**Serial block-face electron microscopy (SBEM) also known as microtome serial sectioning SEM**

Due to the tedious, time-consuming, uncertain, and expensive nature of transmission electron microscopy imaging, the development of an electron microscope with in-built mini ultramicrotome was conceptualized. The idea was to install an ultramicrotome near the specimen chamber of a scanning electron microscope where the samples could be sectioned and imaged under a vacuum. Though the prototype was built (Leighton 1981), the commercialization of such a microscope was achieved much later in early 2000 with the development of serial block-face electron microscopy or SBEM that could process many sections due to automation and capture digital images that can be further processed by associated software. The SBEM consists of a diamond knife mounted on a mini ultramicrotome which is again situated on the inner face of the vacuum chamber door of a scanning electron microscope. The sample is manually moved closer to the diamond knife until it is close enough so that the knife can perform serial sectioning. Once the proper position of the sample block and the knife are adjusted, vacuum is turned on and the parameters for imaging and sectioning are set and serial imaging can be done automatically guided by computer software. On every cycle, the diamond knife sections a portion of the sample of predefined thickness and the exposed smooth surface is imaged by the scanning electron microscope. Thus, many sections can be imaged that can be used to reconstruct the three-dimensional view of the sample using associated software. This technique had a significant impact in biological research where cells, tissues and macromolecular architecture could be viewed in three dimensions which helped in better understanding of biological processes (Calì et al. 2016).

**SEM-FIB focused ion beam SEM (FIB-SEM)**

Since the development of electron microscopy by Knoll and Ruska it has been pivotal for biomedical and fundamental research on biological sciences because of high resolution images that can be obtained by the technology. Even though electron microscopy has been the platform for many biological and biomedical investigations for the last few decades, the long and cumbersome process of sample preparation that is usually done manually leaves a significant space for errors and artifacts. To get volume data, each layer of the sample must be sectioned using an ultramicrotome, stained and loaded in a grid before observing in a transmission electron microscope. Sectioning of samples for transmission electron microscopy is usually done by diamond knife and can be error prone. Besides this, the managing of such volumes of serial sections and preventing loss of any sections during the sectioning process is also difficult to achieve. It is also important to align each section in proper order to reconstruct a three-dimensional image that truly or closely represents the original sample. However, such tedious steps are necessary if we want to capture three dimensional images of cells and organelles along the z axis using transmission electron microscope. Undoubtedly such processes are time consuming, labor-intensive, costly and have a lot of associated factors that cause uncertainty. However, significant improvements in technology in terms of automation and digital imaging in the last few years have resulted in the development of sophisticated scanning electron microscope called FIB-SEM or focused ion beam scanning electron microscope. Such microscopes are easier to use, with more automation thus can yield reliable data in a much shorter time. They are equipped with powerful digital cameras and do not depend on films for image reproduction. They can capture and process large amount of data and are integrated with sophisticated software that can reconstruct three dimensional images of cells to yield detailed structure of the cellular matrix, microtubules, and the biomolecular machinery of cellular metabolism. Thus, several complex questions related to developmental biology, ion transport and neuroscience can be addressed with the assistance of FIB-SEM (Xu et al. 2017).

Unlike confocal microscope which uses lasers to penetrate deep into the cell or tissues, transmission electron microscopes are limited because electrons have very little mass and cannot penetrate deep. Thus, thin sections of samples are made within a range of 70 to 120 nanometers and such thin sections are a prerequisite for imaging using electron microscope. In the case of scanning electron microscopes, thin sectioning is not a requirement, but it yields only topographic data. If a high energy electron beam is used, then the electron penetrates deeper to a few micrometers but results in a substantial amount of noise thus shielding the information of the topography. To circumvent this problem a combination of electron beam and ion beam is used in FIB-SEM. The electron beam is used for imaging while the ion beam is used for milling. Usually, high energy gallium ions are used to create a beam for in-situ milling while the electron beam is used for imaging. The ion beam and automation involved offers more control, higher precision, higher accuracy to the user as compared to diamond knife and much thinner sections of few nanometers thickness can be removed gradually and the exposed face can be imaged simultaneously to yield rich three-dimensional data of samples. Besides milling, FIB-SEM is also used for deposition of materials such as carbon or platinum atom for micromanipulation of samples. The use of FIB-SEM is becoming increasingly common in biomedical science, material science, ceramics, superconductor industry and chemical engineering (Kizilyaprak et al. 2019).

**Cryogenic FIB-SEM**

Biological samples can be frozen to preserve cellular architecture and cryo-FIB-SEM can be applied to obtain high resolution (nanoscale) 3D imaging of biological samples. However, biological samples are made of elements having lower molecular weight thus are not ideal for electron microscopy unless stained with heavy metals. Thus, imaging unstained samples encounters challenges such as low contrast and charging. This problem can be overcome by using cryogenic plasma FIB/SEM to produce high resolution images of biological samples (Dumoux et al. 2023).

**Environmental SEM Environmental/ low vacuum scanning microscopy**

Electron microscopes require vacuum to operate because electrons have very small mass and can easily be deflected by gaseous molecules. If electrons get deflected and cannot be focused, then a high-resolution image cannot be generated. Such an environment causes unique challenges to biological samples. Biological samples are hydrated, and a large portion of their biomass is water. In vacuum water starts to boil and evaporates thus causing severe mechanical damage to the cells and tissues. Thus, the structure of the cells and tissues can be distorted resulting in artifacts. To circumvent this problem most biological samples are dried by using a technique called critical point drying where water in biological samples is replaced by carbon dioxide at critical point which again sublimes thus minimizing the effect of surface tension of water on biological tissues and largely preserving their integrity. However, this process is not completely devoid of artifacts though it helps to preserve sample integrity. A new technological breakthrough in scanning electron microscopy is the environmental scanning electron microscopy where sufficient vacuum can be maintained in the electron column of the microscope while the sample in the specimen chamber can be placed in a gaseous atmosphere. Imaging in the environmental mode is challenging and, in many cases, high resolution images are difficult to obtain, however, this allows imaging of samples that are hydrated. Thus, biological samples can be directly imaged without significant sample preparation (Muscariello et al. 2005).

**Helium Ion Microscopy (HIM)** is a relatively recent development in the field of scanning microscopy where instead of electrons, helium ions are used for imaging. The uniqueness beyond this imaging technology is the source of imaging ions. The ions used for imaging are generated preferentially from three atoms in the apex of the source and only one of them is utilized for imaging. The ions are accelerated down the column of the microscope in a similar fashion as in electron microscope. The beam generated is very bright and thus a very small aperture can be used for imaging which reduced the chances of chromatic and spherical aberration while increasing depth of field. Some other advantages of this microscope are high secondary electron yield which reduces noise and there is no charging of samples, so the non-conductive samples are not coated with gold which contribute to better imaging with a resolution below one nanometer (Hlawacek et al. 2014).

**EDS**

Energy dispersive X-ray spectroscopy or EDS is a semi-quantitative or ratiometric based microanalysis technique which is utilized to understand the composition of materials including biological cells and tissues. It is also known as electron probe X-ray microanalysis or EDX. Most electron microscopes are usually associated with EDS capabilities. Electron microscopes use high energy electron beams that impinges the sample. The electrons interact with atoms of the sample that results in elastic and inelastic scattering of electrons, the electron loses energy in the case of inelastic scattering but does not lose any energy in the case of elastic scattering. Such interactions result in emission of secondary electron, backscattered electron, auger electrons, cathodoluminescence and characteristic X-rays with dissipation of heat (Figure 1). EDS detectors can identify characteristic X-rays and such identification are the basis of element identification and their special resolution within a sample. When atoms in the samples are subjected to high energy electron beam of the electron microscope, electron in their orbital gets excited. However, when these electrons return to their ground state, they emit X-rays. The energy of such X-rays is same as the difference between the energies of the two orbitals involved in the transition. Such energies are characteristic of elements or atomic numbers such that each element has their own characteristic X-ray energy that can be detected by EDS detector. The element detection based on X-rays are represented using a spectrum or histogram where X-ray counts are plotted along with energy to yield peaks. The x axis represents energy of X-rays while the Y axis represents the counts of X-rays. The position of the peak along the X axis (energy) represents the element that is detected while the area of the peak represents the quantitative data or amount of a particular element in a sample. Several integrated software such as ESPRIT family from Bruker, IXRF systems and APEX from Gatan are commonly used for EDS analysis (Allen et al. 2012).

One of the common problems in EDS is the generation of continuous X-rays. It is produced because high energy electrons are slowed by the electrostatic field of the nucleus of the atoms of the sample, this results in a background X-rays that can shield the element peaks to a certain extent and cause difficulty in peak identification and separation. Besides this, the technology cannot distinguish between different valance states of elements thus it detects elements independent of their ionic characteristic. It is also important to understand that like electron microscopes, the EDS technology operates in vacuum as X-rays can be absorbed by air molecules or can be deflected, thus preparation of biological samples for EDS analysis are exceptionally challenging. Though dry material samples such as rocks, ceramics, metals, and minerals are usually easy to analyze by EDS, the biological samples which generally has high water content needs to be dehydrated before they can be used for EDS analysis. The most common method of dehydration is using critical point dryer, but the user needs to be very careful and design protocol that interferes least with the sample to ensure that the chemical composition of the sample remains unchanged representing their native state. EDS is a very useful technology for element identification but is much less sensitive as compared to some other available spectrometric technologies such as ICP-OES. Elements present in parts per million range can be detected by EDS while elements with lower atomic weight can be challenging to detect if they are present in low amount in the sample analyzed and this often results in peak overlapping in case of low atomic weight elements. The spatial resolution of EDS is on the range of few micrometers and accelerating voltage can affect such resolution parameters (Abd Mutalib et al. 2017; Newbury & Ritchie 2013). EDS quantifies elements ratiometrically; thus, it is considered as semi-quantitative in nature, and it is difficult to determine the exact concentration of an element in a sample. However, known concentration of salt solution can be used to make standards for quantification of samples. Biological samples are made of mostly elements with low atomic weight thus are particularly challenging for EDS analysis due to their low generation of characteristic X-rays and formulating standards for them is also cumbersome. However, such standards were made using *Pinus* (Pesacreta & Hasenstein 2018), *Zea* (Pesacreta et al. 2021) and *Arachis* (Acharya & Pesacreta 2022) seedling roots to investigate ion accumulation in seedling root tissues (Figure 2; Figure 3 & Figure 4). In the above-mentioned protocol, the authors used the seedling root itself to act as a matrix for standards. They treated the root samples with liquid nitrogen followed by thawing and finally submerging them in deionized water followed by different standard solutions. The liquid nitrogen treatment, thawing and submersion of the root samples in deionized water allowed the endogenous elements to diffuse out of the root cells while the treatment with known standards allowed the roots to have similar concentration of elements as the respective standards. This methodology gave excellent results in estimating low atomic weight element concentration in biological samples as validated by ICP-OES measurements and can may be utilized for other biological tissues. It is important to understand that the generation of characteristic X-rays is dependent on several factors besides the elemental composition of samples. The accelerating voltage, filament saturation level, energy of electrons, tilt of stage, coating of samples, vacuum level are the few factors among many that can influence X-ray signals. To keep all factors that influence the generation of characteristic X-rays in an electron microscope is theoretically possible but in practicality it is extremely difficult. Due to such limitations, the X-ray count obtained for the same sample might differ if measured at different time points, even though the microscope operation parameters are kept same. This problem is more pronounced when elements of low atomic numbers are measured. To circumvent this problem, the protocol for EDS quantification using root standards as described above was improved while estimating ion accumulation in *Zea* (Pesacreta et al. 2021) seedling roots. To have more reliable X-ray count data, the same stainless-steel stub was used as a sample holder for all measurements across different time points. The rim of the stub was used to procure X-ray counts for iron (Fe) before any samples were investigated using EDS. The Fe counts were used as reference to normalize data of the X-ray counts for other elements in the sample. Such an innovation reduced the errors that can be caused by differential emission of X-ray signals due to microscope parameters that are not within the control of the user. The robustness of the data was supported by ICP-OES analysis of root samples.

EDS is a versatile and highly effective technique for element mapping and identification, thus is used in biological, biomedical, engineering, nanotechnology, and material science. It is widely used to identify nanoparticles in biological systems related to chemotherapy and drug delivery. Heavy metals are a serious threat to human and animal health and cause significant damage to the environment. Though some heavy metals are required in very minute amount in the biological systems, however most of them are not necessary and their accumulation causes significant physiological and medical conditions related to cancer, mutagenesis, endocrinal and neurological diseases. The main source of heavy metals is industrial waste, vehicle emission, military equipment, agricultural fertilizers, herbicides, and pesticides. EDS is widely used for the detection of heavy metals in environmental samples, crop plants and within cells and tissues of living organisms. The bioaccumulation of heavy metals, tissue calcification and accumulation of minerals in cells and tissues are related to several diseases and EDS offers a quick and inexpensive method to detect these anomalies in biological system. Nanoparticles are defined as particles having a size of less than 100 nm and are an increasingly important area of research due to their physical and chemical properties. Recently, scientists in the field of nanotechnology are moving away from chemical synthesis and are increasingly using biological extracts or systems to synthesize nanoparticles for environmental, financial, and medical reasons. Thus, the application of nanotechnology has overlapped most scientific disciplines from agriculture, health to product development that are of commercial importance. The response of biological systems and molecules to nanoparticles is an important area of research and in this context the use of silver and gold nanoparticles are particularly important. Significant amount of scientific information regarding the synthesis, structure and function of nanoparticles is available due to the use of EDS and electron microscopy, thus EDS had a pivotal role in the advancement of nanotechnology (Wyroba et al. 2015; Mishra et al. 2017).

**Application of SEM/EDS**

**Plant science research**

SEM/EDS is widely used to understand several physiological processes in plants. A multitude of investigations involving cryo-fixed samples of plants resulted in significant understanding of mechanisms related to salt tolerance, biomineralization, phytoremediation, root pressure and guttation. Scanning electron microscope coupled with energy dispersive spectroscopy is widely used in the physiological studies involving hyperaccumulators (Dong et al. 2019). The source-sink relationships, ion uptake by roots and element compartmentation in plant has significant implications in agriculture and environment. The advancement of SEM/EDS techniques coupled with cryo-techniques has enabled plant scientists to investigate element localization at a cellular and tissue level. Cryo-fixation involves rapid freezing of hydrated plant samples and the water sublimes; thus, the effect of surface tension can be reduced preventing damage to hydrated plant structures and preserving their structural integrity. Mycorrhizal fungi are essential for the health of plant, they associate with plant roots as ectomycorrhizae or endomycorrhizae and benefits the plants through sequestration of nutrients such as phosphorous and acts as a below-ground communication network between distant plants while the plant provides photosynthates to the heterotrophic fungi. Insights into the mutualistic relationship between plants roots and the mycorrhizae was possible due to SEM/EDS analysis of such associations (Wu et al. 2015). A large amount of research has been performed to understand the uptake of ions from roots to the aboveground parts of plants while scientific literature regarding the translocation of solutes and ions from seeds to seedling root tip is not well defined. Most of the ion transport studies involved radioactive compounds or xenobiotic dyes and compounds which can have detrimental effect on plant physiology and can negatively influence the results of ion transport studies in plants. To circumvent this problem and to study the transport and distribution of endogenous ions from seeds to seedling roots, cryo-SEM/EDS was used. Seeds from several distantly related species were germinated using nanopure water to minimize the influence of exogenous ions and the ion transport and distribution were investigated using cryo-fractured cross section of seedling roots observed and analyzed using SEM/EDS (Pesacreta & Hasenstein 2018; Pesacreta et al. 2021;

Acharya 2021; Acharya & Pesacreta 2022; Acharya & Pesacreta 2023). Plants are composed of elements of low molecular weight and thus low concentration of such elements in plant cells and tissues poses challenges for detection by the EDS detector. Though theoretically EDS has a detection threshold of parts per thousand, a common challenge is peak overlapping of elements which have close X-ray energies or shielding of characteristic peak by continuous X-rays. The spatial resolution of such analysis can be around 3 micrometers (Goldstein et al. 2017).

**Nanobiotechnological applications.**

Richard Feynman in 1959 coined the term nanotechnology that involves the synthesis and modification of materials at a nanoscale. As the size of an object decreases, its surface to volume ratio increases. This property of particles imparts unique characters and functionality to them. When the particles reach a dimension of 100 nanometers or less then they are classified as nanoparticles. Apart from genomics, very few research areas have attracted the attention of scientists and the public in general as compared to nanoparticles in the last few decades (Figure 5). The considerable interest in the topic is due to the novel physical, chemical and optical properties of nanoparticles and their potential application in medical, environmental, forensic, and material sciences. This has resulted in the development of a relatively new area of science called nanobiotechnology. Catalyst plays a significant role in chemical and biochemical reactions thus the nanoparticles with catalytic properties are of great importance and thus the elucidation of structure and composition of catalytic nanoparticles are an important research area. Several factors in the chemical synthesis of nanoparticles such as temperature, pressure, pH, and the chemical environment influences the fundamental properties of nanoparticles such as size, orientation, structure, clustering, stability, and the overall efficiency of the synthesis process. Though significant progress has been made in the chemical synthesis of nanoparticles, they are expensive, time consuming, inefficient, environmentally harmful and in many cases cannot be regulated thus yielding undesirable results. Chemically synthesized nanoparticles might also have negative immune response and can be rejected by human immune cells thus limiting their applications. Use of biological organisms such as microorganisms, algae, and plants or biomolecules extracted from them such as enzymes and proteins in the synthesis of nanoparticles is termed as "green synthesis". The green synthesis of nanoparticles offers significant benefit over chemical synthesis methods by being more cost effective, environmentally friendly, and usually these are a one step process that can be conducted over a normal physiological temperature, pressure, and pH and with significant control and efficiency while yielding stable and consistent results thereby improving efficiency process. Thus, green synthesis of nanoparticles or synthesis of nanoparticles utilizing biological metabolic pathways is gaining momentum and in near future might accelerate targeted drug delivery to treat several diseases including cancer. Several different types of nanoparticles such as gold nanoparticles, silver nanoparticles, carbon nanotubes and organic nanoparticles are widely synthesized, and their characteristics are quantified. To understand the characteristics of nanoparticles and to relate such characteristics to structure and composition of nanoparticles, it is essential to visualize them at a high resolution followed by compositional analysis at high spatial resolution. Instruments such as UV-vis spectroscopy, transmission electron microscope, atomic force microscope and Raman microscope, X-ray diffraction and nuclear magnetic resonance are widely used for the process, however, the used of scanning electron microscope along with energy dispersive spectroscopic analysis offers a relatively fast and dependable method to visualize and characterize nanoparticles. (Patra & Baek 2015; Vida-Simiti et al. 2004; Chen 2011; Shao et al. 2011).

**Application of SEM/EDS in biological research**

Microscopes played a major role in understanding cellular structure and processes of pathogens causing diseases. The advent of electron microscopes accelerated such investigations because it allowed us to study cells with greater details, however, sample preparation remained a major bottle neck. The improvement of technology including development of protocols to preserve samples such as critical point drying, immunolabeling and freeze-substitution allowed better resolution and reduced artifacts thus making it possible for higher resolution imaging of cellular structures. Other areas of improvement included improved electromagnetic lenses and more sensitive detectors which accelerated biomedical research including the investigation of pathogenicity, host-parasite interactions, and cellular response of several parasites such as Toxoplasma, Giardia and Trichomonas (de Souza & Attias 2018).

Due to anthropogenic effect and climate change several ecosystems have been disturbed including the very sensitive and important coral reef ecosystem. The changing pH of marine habitat has caused damage to ecosystems that impact the flora and fauna of such habitats. Scanning electron microscopy along with energy dispersive spectroscopy has been a useful tool to study the changing coral ecosystem and investigate the diversity of marine life (Arreola-Alarcón et al. 2022). The use of heavy metals in several industrial processes has rendered many agricultural and arable land unusable and toxic. Phytoremediation is an inexpensive, ecofriendly, and sustainable method of cleaning heavy metal contaminated soil and analytical techniques such as SEM/EDS has been a useful tool to investigate the mechanism of phytoremediation and the physiology of several plants that can hyperaccumulate metals (Jesitha et al. 2018). SEM is widely used in the identification of different insect species. Insects are the largest group of organisms in the animal kingdom, and they are extremely important for our ecosystem. Many insects are also very damaging to agricultural crops and are considered pathogens while bees are important for pollination of crops and thus very relevant for agriculture and food security. Since insects are diverse and very large in numbers, their study can lead to better understanding of evolution and the diversity of life. For a multitude of reasons, insects are widely studied, and SEM is frequently used to investigate their morphology and identify for agricultural, ecological, and evolutionary studies (Watts et al. 2022). Another significant application of SEM is in the study of pollens or spores often termed palynology. Since pollens have agricultural, archaeological, geological, and environmental significance, the study of pollens is a broad area of research. The pollen can be of different size, shape and may have different architectures that can be visualized clearly using a scanning electron microscope. The technique is commonly used to identify pollens and their dispersal (Bahadur et al. 2018). SEM has also been used in the study of geckos and to understand the morphological and physical mechanisms they employ to move or adhere to walls, rocks, or other objects (Bonfitto et al. 2022). SEM is widely used in biomedical research to study the effects of microgravity in living systems and to understand biofilm formation by bacteria (Kallapur et al. 20021). Thus, the application of SEM/EDS is spread over a range of biological and biomedical studies. Since the SEM/EDS technique is versatile, powerful, and relatively easy to use, it has been applied in various areas of biological and biomedical science.

**Conclusion**

Since the development and commercialization of scanning electron microscopes, it has been used widely in biological sciences, biomedical sciences, material sciences and engineering. The versatility and relative ease of use of this instrument has made it a workhorse of different scientific research across the globe. Improvements in filaments and electron guns such as cold emission along with development of powerful electromagnetic lenses and detectors has resulted in increasing the resolving power of the microscope. With the advent of FIB-SEM, three dimensional images of materials and cells are possible thus adding more power to the microscope which could eventually replace transmission electron microscopes. The FIB-SEM allows reliable imaging along the z-axis and has high z-resolution. Automation and powerful software for image analysis has made this instrument a fast, reliable, and high throughput tool for acquiring three dimensional images of biological and material samples. However, there are many challenges that need to be addressed. Vacuum is one of the major bottlenecks that limits the electron microscope imaging of hydrated biological samples. Since most biological samples have a high amount of water, it is essential to remove the water using a critical point dryer before placing the sample in the specimen chamber of the microscope. Another major problem that biologists face while imaging biological sample is their inability to conduct electrons through the samples thus the electrons from primary beam accumulate resulting in charging. It creates a very bright image thus shielding the topographic features of the sample. The development of environmental or variable pressure scanning electron microscopes has allowed imaging of hydrated samples, however, the technology is still at its infancy and needs substantial improvement for high resolution imaging of hydrated samples. The future of scanning electron microscopy is exciting because more powerful microscopes are being developed by manufacturing companies such as JEOL, Hitachi, FEI and Thermo Fisher along with increase in the use of this technology in a wide area of research.

**Acknowledgement:** I thank Delta State University for supporting my research endeavor

**Conflict of Interest:** The author declares no conflict of interest.

**Funding:** The author received no funding from any source for this work.

**Figures:**

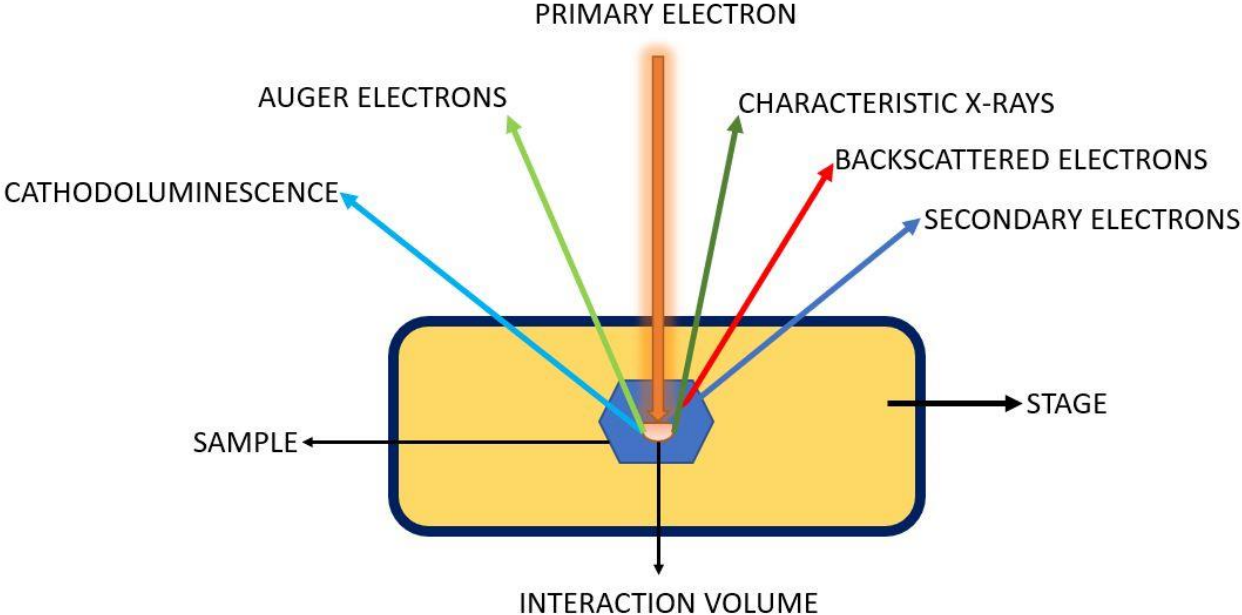

Figure 1. Electromagnetic interactions between primary beam and sample

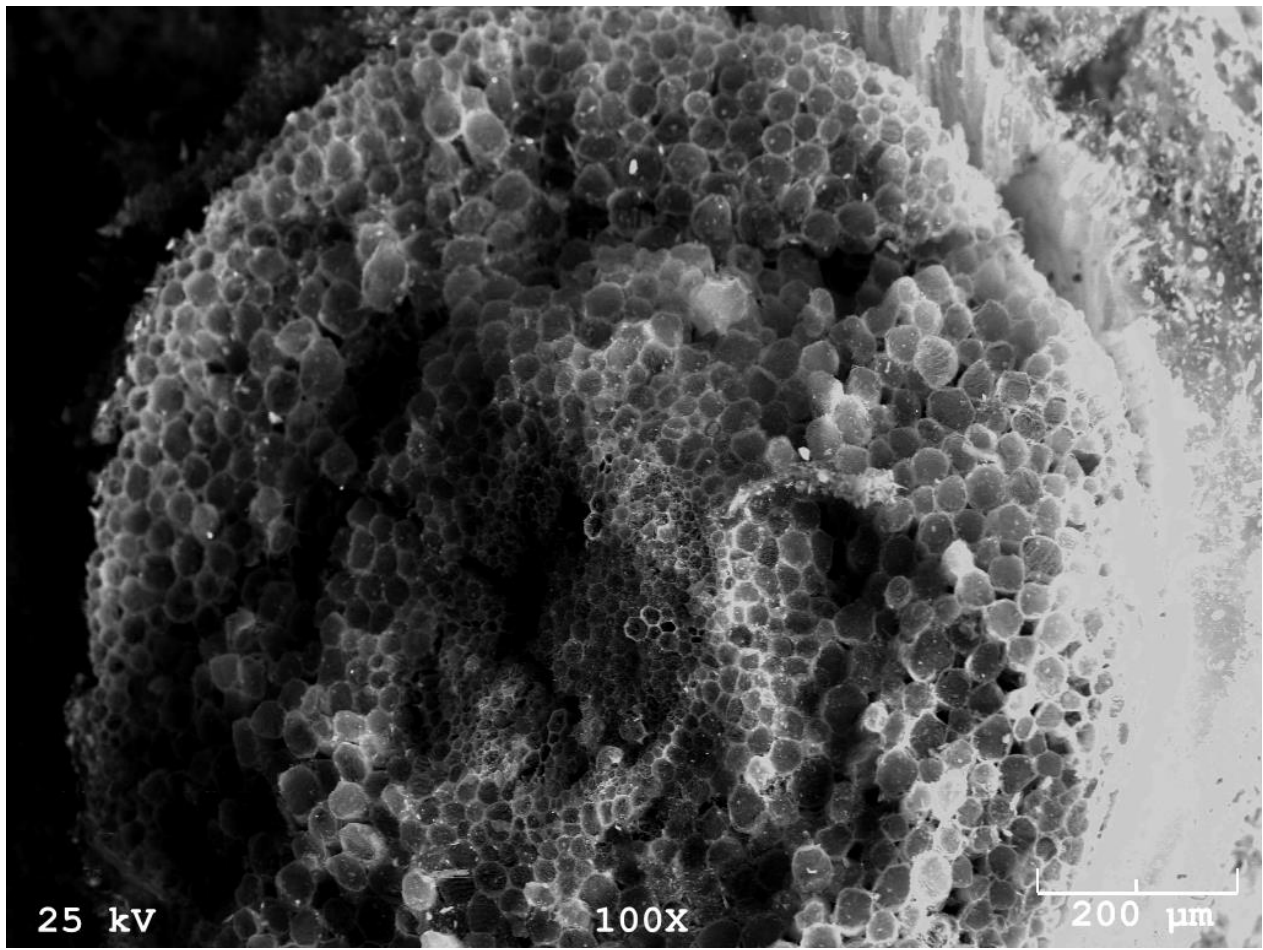

Figure 2. A backscattered image of cryo-fractured *Arachis hypogaea* seedling root at 15 mm from the root tip.

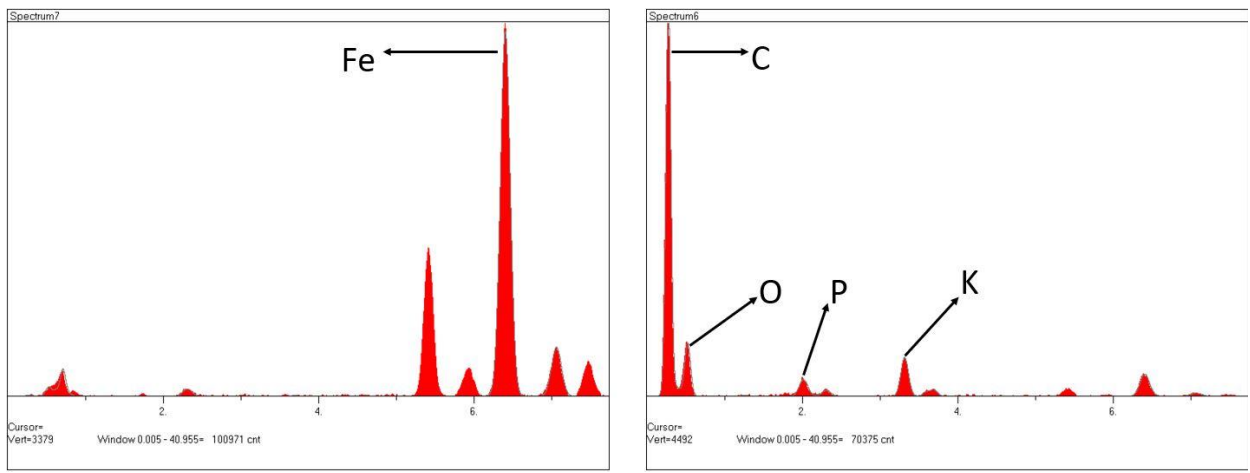

Figure 3. The EDS spectrum showing Fe peak from the stub and C, O, P and K peak from the seedling root sample.

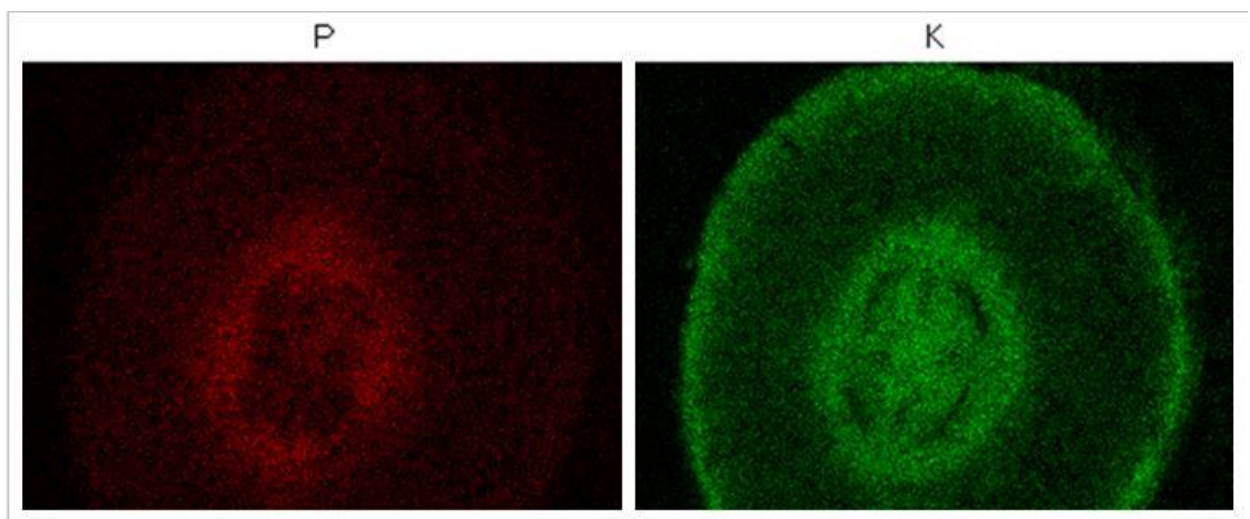

Figure 4. Images of elemental mapping represent K and P distribution in the seedling root of *Arachis hypogaea* at 15 mm from the root tip.

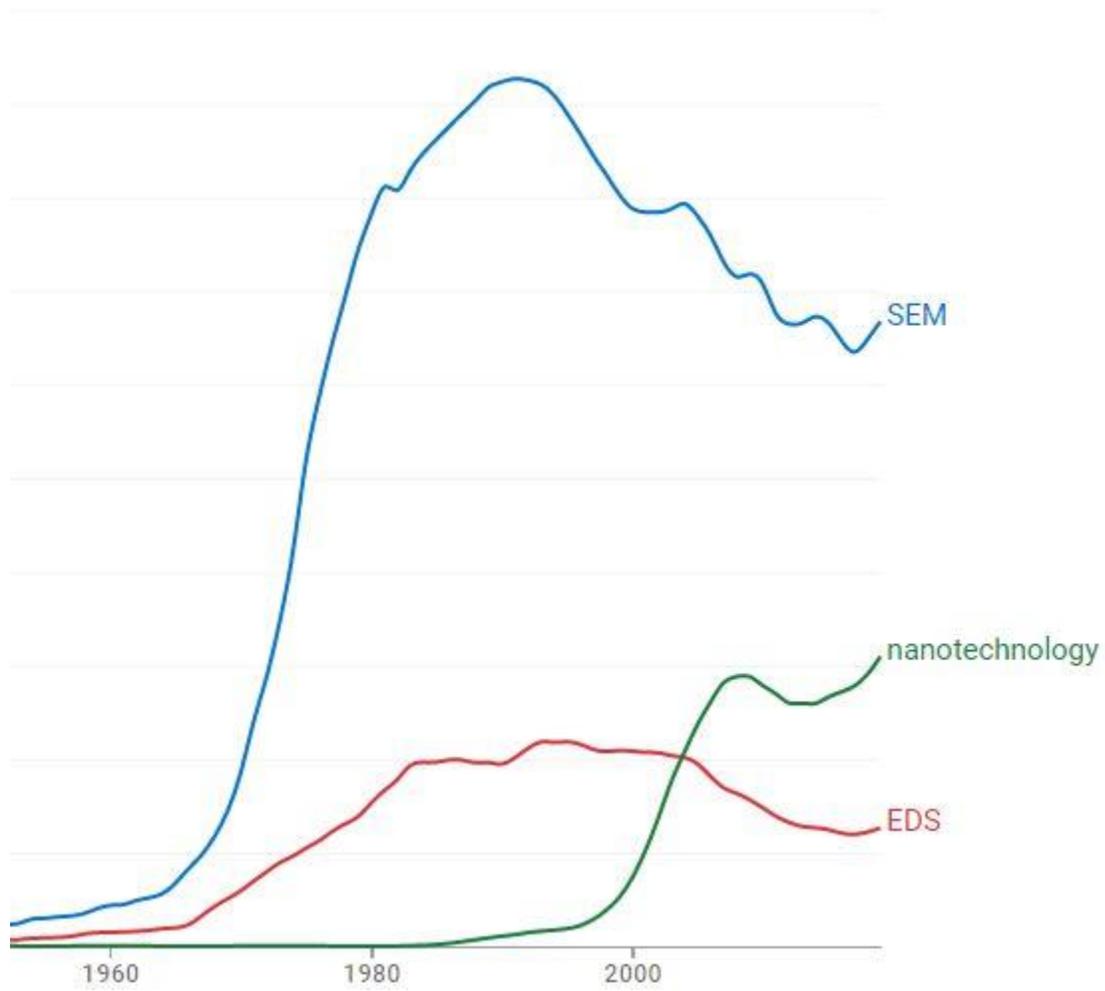

Figure 5. A Google Ngram graph representing the relative use of the terms SEM, EDS, and nanotechnology from 1960 to 2019.

**Table:**

| | | |
|---|---|---|
| 1590 | Compound Microscope built by Janssen | Chandler & Roberson 2009 |
| 1870 | Abbe achieves a resolution of 250 nm | Chandler & Roberson 2009 |
| 1897 | Electrons discovered by Thompson | Chandler & Roberson 2009 |
| 1910 | Microscopes coupled with camera | Chandler & Roberson 2009 |
| 1926 | Deflection of electrons by magnetic lens shown by Busch | Chandler & Roberson 2009 |
| 1931 | Electron microscope built by Knoll and Ruska | Chandler & Roberson 2009 |
| 1935 | Phase-contrast microscope built by Zernicke | Chandler & Roberson 2009 |
| 1938 | Scanning electron microscope was built | Chandler & Roberson 2009 |
| 1960 | Everhart and Thornley developed secondary electron detector | Goldstein et al. 2003 |
| 1960 | Energy dispersive spectroscopy developed | Chandler & Roberson 2009 |

| 1965 | Scanning electron microscope was available commercially | Bogner et al. 2007 |
| --- | --- | --- |
| 1966 | SEM used for K and Cl localization in Corn | Läuchli & Schwander 1966 |
| 1980 | Focused ion beam development | Giannuzzi 2004 |
| 1980 | Development of Environmental SEM | Stokes 2003 |
| 1982 | Scanning tunneling microscopy developed | Chandler & Roberson 2009 |
| 1985 | Confocal microscope became commercially available | Chandler & Roberson 2009 |
| 1990 | CCD cameras developed | Chandler & Roberson 2009 |
| 2004 | STEM imaging with a resolution less than 1 Angstrom | Nellist et al. 2004 |
| 2004 | Serial Black Face Scanning electron microscopy | Denk & Horstmann 2004 |
| 2006 | Three-dimensional imaging of biological specimen using FIB-SEM | Heymann et al. 2006 |
| 2006 | Development of ATUM-SEM | Hayworth et al. 2006 |
| 2018 | Electron microscopy captures images of less than 0.5 Angstrom resolution | Jiang et al. 2018 |

Table 1. The table illustrates the chronological development of scanning electron microscope

**Figure Legends:**

Figure 1. Electromagnetic interactions between primary beam and sample

Figure 2. A backscattered image of cryo-fractured Arachis hypogaea seedling root at 15 mm from the root tip.

Figure 3. The EDS spectrum shows Fe peak from the stub and C, O, P and K peak from the seedling root sample.

Figure 4. Images of elemental mapping represent K and P distribution in the seedling root of Arachis hypogaea at 15 mm from the root tip.

Figure 5. A Google Ngram graph representing the relative use of the terms SEM, EDS and nanotechnology from 1960 to 2019.